# Interaction of counter-streaming plasma flows in dipole magnetic field

**I F Shaikhislamov, V G Posukh, A V Melekhov, P A Prokopov, E L Boyarintsev, Yu P Zakharov and A G Ponomarenko**

Dep. of Laser Plasma, Institute of Laser Physics SB RAS, pr. Lavrentyeva 13/3, Novosibirsk, 630090, Russia,

e-mail: ildars@ngs.ru

*Abstract:* Transient interaction of counter-streaming super-sonic plasma flows in dipole magnetic dipole is studied in laboratory experiment. First quasi-stationary flow is produced by Ө-pinch and forms a magnetosphere around the magnetic dipole while laser beams focused at the surface of the dipole cover launch second explosive plasma expanding from inner dipole region outward. Laser plasma is energetic enough to disrupt magnetic field and to sweep through the background plasma for large distances. Probe measurements showed that far from the initially formed magnetosphere laser plasma carries within itself a magnetic field of the same direction but order of magnitude larger in value than the vacuum dipole field at considered distances. Because no compression of magnetic field at the front of laser plasma was observed, the realized interaction is different from previous experiments and theoretical models of laser plasma expansion into uniform magnetized background. It was deduced based on the obtained data that laser plasma while expanding through inner magnetosphere picks up a magnetized shell formed by background plasma and carries it for large distances beyond previously existing magnetosphere.

## 1. Introduction

Laboratory modeling of space plasma processes is an important method of study of basic physics. Despite of significant progress in spacecraft measurements and numerical simulations a laboratory experiment remains a source of unique data inaccessible by other means. One of the fields where namely laboratory experiments have pushed the advances in theory and numerical simulations is interaction of counter-streaming plasma flows in presence of magnetic field. In 1970-s and 80-s a number of works with laser-produced plasma expanding with super-Alfvenic velocity into magnetized background have been carried out (*Paul et al. 1971, Cheung et al. 1973, Borovsky et al. 1984, Antonov*

*et al. 1985*) with the aim to model active near-Earth releases AMPTE, CRRES, Argus, Starfish. Based on obtained results a new dynamic model of interaction – magnetic laminar mechanism (*Golubev et al. 1978*) or finite Larmor coupling (*Winske and Gary 2007*) – has been developed which supplemented the earlier kinematic model of displaced electrons (*Longmire 1963, Wright 1971*). Our recent experiments (*Zakharov et al. 2013, Shaikhislamov et al. 2015a*) provided detailed data verifying both of the models.

The other field which was extensively studied by means of laboratory experiment is magnetosphere (*Podgornyi and Sagdeev 1966*). At KI-1 simulation Facility such studies are based on two sources of plasma – induction Θ-pinch and laser plasma (LP) – which interact with compact magnetic dipoles *(Ponomarenko et al. 2001, 2004)*. Combination of energetically and spatially different plasma flows allowed modeling of extreme compression of the Earth's magnetosphere by super powerful CME or by artificial near-Earth releases (*Ponomarenko et al. 2008, Zakharov et al. 2007, 2008*). Such complex system as field-aligned currents connecting boundary layer with ionosphere has been studied in detail (*Shaikhislamov et al. 2009, 2011*). In the latest experiment a pulse of plasma with a frozen-in transverse magnetic field interacting with magnetosphere has been modeled (*Shaikhislamov et al 2014a*). The flow with transverse frozen-in field has been generated by means of laser-produced plasma cross-field expansion into background plasma which fills vacuum chamber along externally applied magnetic field prior to interaction. The related subject intensively studied at KI-1 Facility is a mini-magnetosphere which can be found above Lunar magnetic anomalies or possibly around magnetized asteroids. Namely laboratory experiments supplied necessary data to formulate and verify a Hall model (*Shaikhislamov et al. 2013, 2014b, 2015b*) which explains unusual features of mini-magnetosphere observed in earlier numerical simulations (*Omidi et al. 2002, Blanco-Cano et al. 2003*).

In the present experiment we investigate essentially new combination of interacting flows and magnetic field. Θ-pinch plasma fills the vacuum chamber and creates around magnetic dipole a magnetosphere with estimated size of about 30 cm. The novel feature is that laser plasma is generated inside of this magnetosphere at two targets symmetrically placed at dipole cover. LP is directed opposite to the Θ-pinch flow and has kinetic energy large enough to sweep previously existing plasma and dipole magnetic field. We study the interaction at distances 40÷90 cm from the dipole beyond the previously existing magnetosphere. It was found that the LP flowing through background plasma partially expels it due to Coulomb collisions, and carries along a magnetic field which is order of magnitude larger than the vacuum dipole field value at these distances. Obtained data suggest a novel and unexpected feature that LP captures magnetospheric field rather than simply stretching it and that effectiveness of this capture is directly related to the density of background plasma which creates magnetosphere.

The specific case of plasma expansion from the inner region of magnetic dipole outward into the background flow has at least two possible applications. It directly relaters to the concept of magnetosail (*Winglee et al. 2000*), which was extensively studied theoretically (*Khazanov et al. 2005*), numerically (for example, *Moritaka et al. 2010*) and in laboratory experiments as well (*Funaki et al. 2007, Slough et al. 2010, Antonov et al. 2013*). The other field is Hot Jupiters – close orbiting exoplanets heated by ionizing stellar radiation to a point of super-sonic expansion of upper atmosphere (for example, *Shaikhislamov et al. 2014c*). The interaction of expanding planetary flow with counter-streaming stellar plasma in case when such a planet possesses weak magnetic field (*Khodachenko et al. 2015*) was one of motivations of the present experiment.

The paper consists of two sections on the experimental set-up and results, followed by the discussion and conclusions.

## 2. Experimental set up and results

Experiment has been carried out at KI-1 space simulation Facility, which includes chamber 500 cm in length and 120 cm in diameter operating at a base pressure of $10^{-6}$ Torr. Induction Θ-pinch with exit aperture of Ø20 cm ejects ionized hydrogen plasma which, for conditions of present experiment, expands and propagates with a velocity of $20 \div 60 \, \text{km/s}$ along the chamber axis and is sustained for duration of about $100 \, \mu s$. At the chamber axis a magnetic dipole was placed at a distance of 310 cm from the Θ-pinch exit aperture. Magnetic moment with a value of $\mu=7.5 \cdot 10^5$ G·cm$^3$ and a fall off time $250 \, \mu s$ was oriented perpendicular to the chamber axis. The dipole has an epoxy cover in form of a cylinder with a size of 5 cm on which a polyethylene target was fasten. Two $CO_2$ laser beams of 70 ns duration and of 150 J energy each are focused symmetrically into spots with a size of about 2 cm. The vectors normal to the surface planes along which the plasma plumes expands were slightly divergent relative to each other. Experimental set up is shown in a snapshot of plasma generated by laser beams in presence of dipole magnetic field (fig. 1) where all elements and the used GSM frame are indicated.

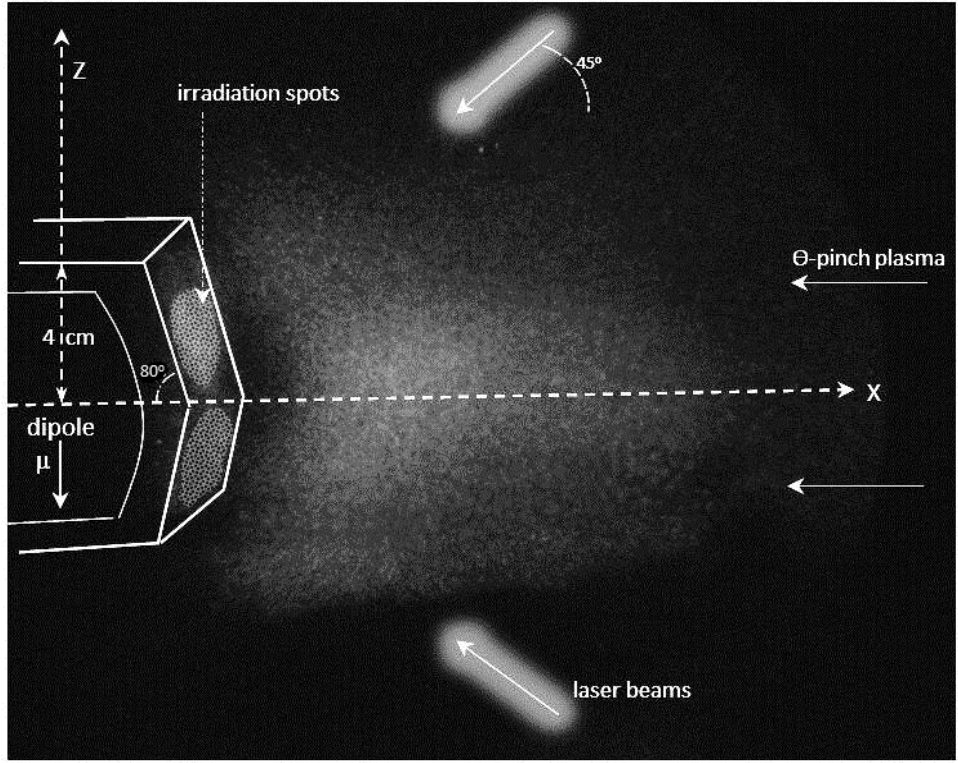

**Figure 1.** Experimental set-up superimposed on snapshot of laser-produced plasma.

LP plasma consisted mostly of $H^+$ and $C^{4+}$ ions approximately in equal parts and each plume expanded inertially with a front velocity of $170$ km/s in a cone with half-angle of about $30°$. A total number of electrons per solid unit angle of the both plumes measured far from the target was about $N_{e,\Omega} \approx 2.2 \cdot 10^{18}$ sr$^{-1}$ and kinetic energy of ions $W_\Omega \approx 32$ J/sr. Due to specific pulse and tail generation mode of laser oscillator there was, besides the first plasma bunch, a secondary and a third flows generated 550 ns and 1100 ns later. The second flow was largest in amplitude and twice as slow in comparison to the first one. Diagnostics consisted of miniature Langmuir and three-component magnetic probes with spatial resolution better than 0.5 cm. There were also directional ion collectors operating similar to a Faraday cup. More detailed description can be found in our previous papers (*Ponomarenko et al. 2004*, *Shaikhislamov et al. 2014a*). Most electric and magnetic measurements were made at distances 40-90 cm from the target. There were also Langmuir probe and ion collector measurements at a distance of 90 cm from the Θ-pinch aperture corresponding to a distance of X=207 cm from the target.

Fig. 2 shows typical probe measurements of Θ-pinch plasma obtained close and far from the Θ-pinch aperture. The flow has sharp switch and decay fronts. Calculated velocities of these and other telling points in both dynamic signals are shown by crosses at the same plot. Velocity profile is aligned with the ion current measured at X=67 cm so the first and the last crosses correspond to the switch and the decay fronts. One can deduce that at the time when laser plasma is created (if laser operates at a time zero) the first front of the Θ-pinch flow moving with velocity of 60 km/s reaches the point X=−300 cm, that is very far downstream of the dipole and laser target. Meanwhile the decay front moving with velocity of about 25 km/s is positioned at this time at X=150 cm. Thus, the Θ-pinch flow creates around the magnetic dipole quasistationary magnetosphere, and the laser-produced plasma which starts to expand at t=0 in the opposite direction interacts with plasma column of 150 cm length. Because of expansion the density decreases with distance from the Θ-pinch plasma origin and corresponding measurements (fig. 3) show that it is close to expected cubic fall off. Note that the density in the region of interest (40-90 cm) is about $n_e \approx (3 \div 5) \cdot 10^{12} cm^{-3}$. Further on the Θ-pinch plasma is called, when convenient, the background plasma.

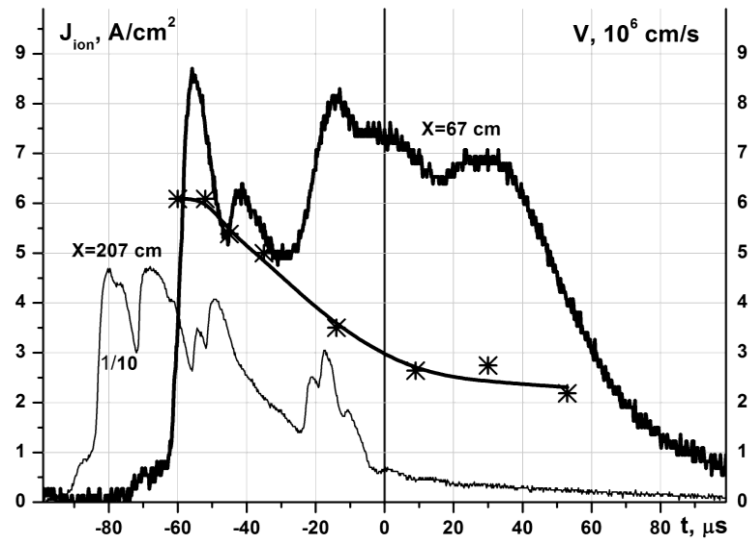

**Figure 2.** Ion current measured in Θ-pinch plasma at two positions along the flow direction (X axis). Position X=207 cm is at a distance of 90 cm from the Θ-pinch exit aperture. Note that the corresponding signal (thin line) is attenuated by 10 times. Position X=67 cm (signal presented by thick line) is in the central region where most other measurements described in the paper were made. Zero time corresponds to the moment when laser beam, when operated, hits the target. Right abscissa is for velocity presented by crosses.

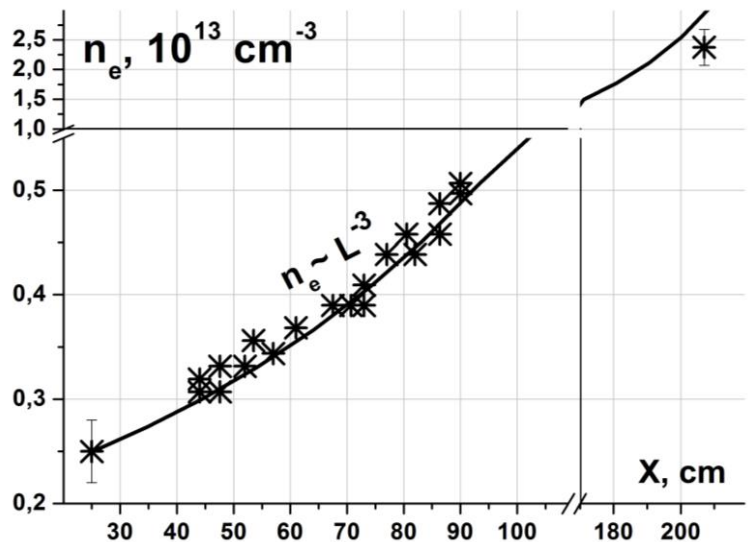

**Figure 3.** Spatial profile of Θ-pinch plasma along the direction of propagation. Solid line shows cubic fall off calculated in terms of distance from the origin point of the Θ-pinch plasma L=330-X.

Laser-produced plasma flow is demonstrated in the fig. 4. The second maximum in the flow has a sharp front, is the largest by amplitude and produces most strong interaction with the Θ-pinch plasma. Such interaction is demonstrated in the next fig. 5. Probe signals show dynamically as background plasma fills the chamber and then the passage of counter-streaming laser plasma. Comparison of signals of directional ion collector and Langmuir probe reveals that the second LP flow effectively sweeps the background significantly decreasing its density.

Despite the expulsion of background the interaction is rather weak because laser plasma practically isn't affected. First of all, it isn't decelerated even after passing large distances through background. This is shown in time-of-flight diagrams (fig. 6) plotted for the fronts of the first and the second LP flows. Second, dynamic signals of LP measured at X=67 cm in vacuum, with added dipole magnetic field and with background plasma as well show little difference (fig. 7). The significant difference was seen only at X=207 cm. The amplitude of the second LP flow was observed to decrease 3-fold, while the first LP flow was affected in a much less degree. Note that the attenuation of the second LP flow depends only on the background plasma regardless of dipole magnetic field which proves that the interaction takes place due to the Coulomb collisions.

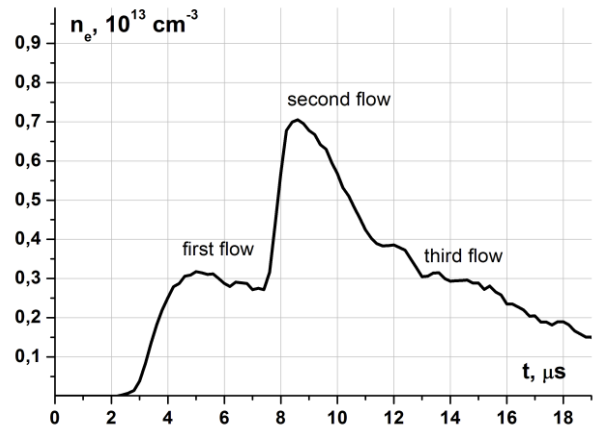

**Figure 4.** Typical dynamic signal of laser-produced plasma measured by Langmuir probe at a distance of X=67 cm from the target. Laser beam hits the target at a moment of t=0. One can see the first, the second and the third plasma flows generated by pike and tail of the laser pulse.

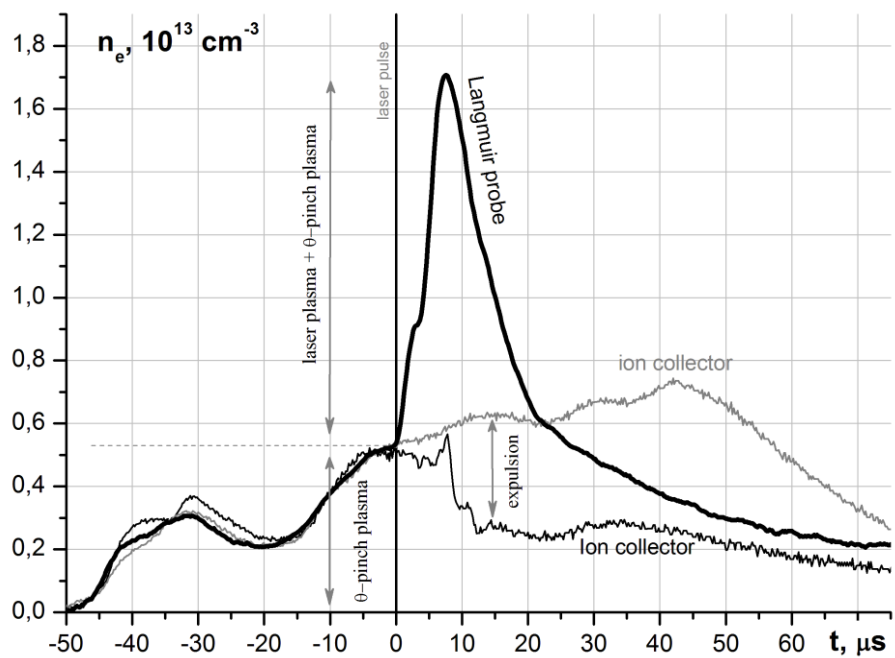

**Figure 5.** Demonstration of interaction of laser-produced and Θ-pinch plasmas. Ion directional collector is oriented towards the Θ-pinch and doesn't register the laser plasma. Thin grey line shows the electron density of Θ-pinch plasma without the laser plasma. When laser operates the ion collector shows partial expulsion of Θ-pinch plasma by laser plasma (thin black line). The Langmuir probe (thick solid line) measures both plasma flows. One can see the laser plasma pulse superimposed on the Θ-pinch signal. Timing shows that the expulsion of the Θ-pinch plasma is produced by the second LP flow, while the effect of the first LP flow is weak. Measurements are obtained at a distance of X=67 cm.

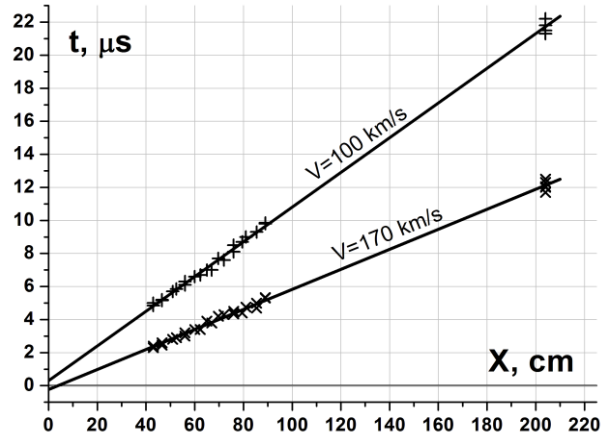

**Figure 6.** Time-of-flight plots t(X) for the front of the first (×) and the second (+) plasma flows.

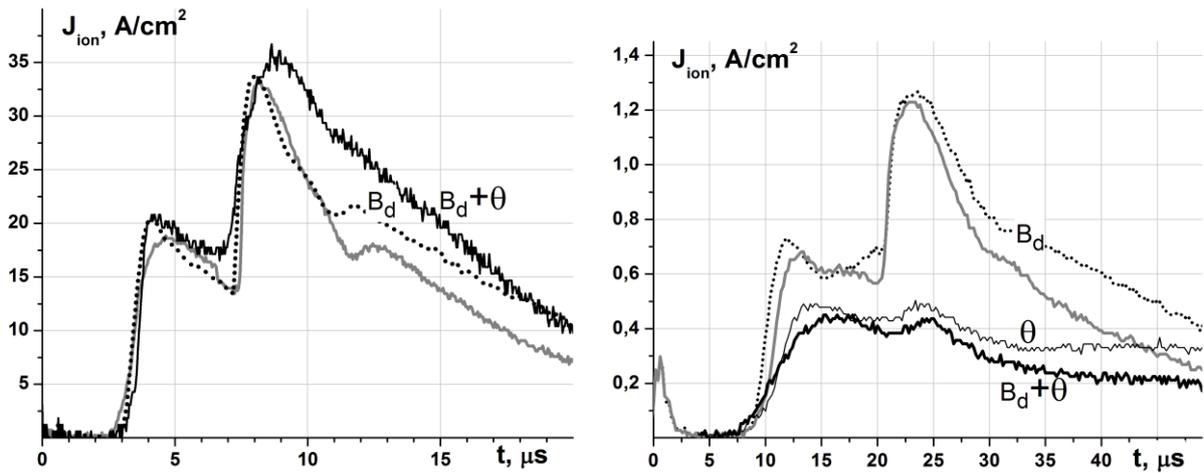

**Figure 7.** Ion current measured at a distance of X=67 (left) and X=207 cm (right) in cases when only laser plasma is present (grey line), when also the magnetic dipole is switched on (dotted, marked $B_d$) and when Θ-pinch plasma is added as well (solid thick black, marked $B_d+\Theta$). The right panel also shows the case when there is Θ-pinch plasma but no dipole field (solid thin black, marked Θ). The measurements at X=67 (left) were performed by Langmuir probe so the ion current generated by Θ-pinch plasma prior to LP arrival has been deducted from the signal $B_d+\Theta$ for better comparison with other signals.

To summarize very briefly, presented so far data show weak collisional interaction of counter-streaming plasma flows which leads to partial sweeping of background. Next we present magnetic field measurements. They mostly concern the largest component collinear to the dipole moment, which is Z-component in equatorial plane. Fig. 8 shows its dynamics in three cases. First one is when LP expands into background plasma but dipole field is switched off. At the foremost LP front a weak and short lived signal of about 2 G was measured. This could be small irregular magnetic field carried by Θ-pinch plasma and swept by laser plasma. When dipole is switched on but the background plasma is absent the LP carries with itself a field which is of the same sign and about the same value as undisturbed dipole magnetic field $B_d = \mu/X^3$ at corresponding distance. The last signal in fig. 8 was measured when both dipole field and background plasma are present. One can see that LP passage brings magnetic field which is of the same direction but much larger in amplitude. Generation by counter-streaming plasma flows of significant transverse magnetic field far from the dipole is a novel finding of the reported experiment.

It should be noted that sufficiently far from the dipole the Θ-pinch plasma expels the dipole magnetic field if it is present. This was observed in our earlier experiments as well in the present one up to the closest distance from the dipole X=44 cm where magnetic measurements were made. Thus, at distances X=40÷90 cm the LP expands in unmagnetized background in which initial magnetic field is virtually zero. Therefore, the observed fields are brought by LP from a region close to the dipole where the initial magnetic field prior to LP generation isn't zero.

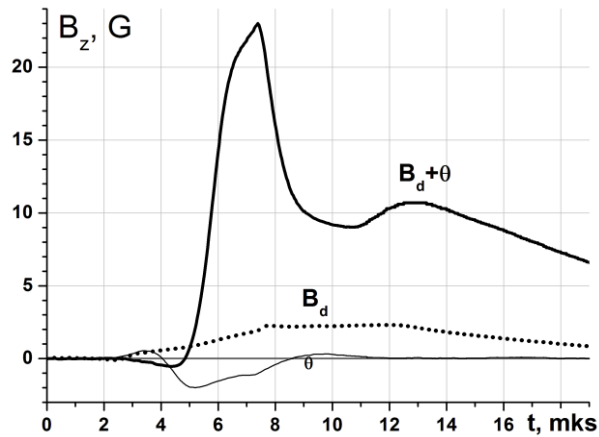

**Figure 8.** Dynamics of the main component of magnetic field measured by probe when laser produced plasma interacts with Θ-pinch plasma without dipole magnetic field (thin solid line), when it interacts with dipole field without Θ-pinch plasma (dotted), and when both Θ-pinch plasma and dipole field are present (thick solid). Probe is positioned at X=67 cm.

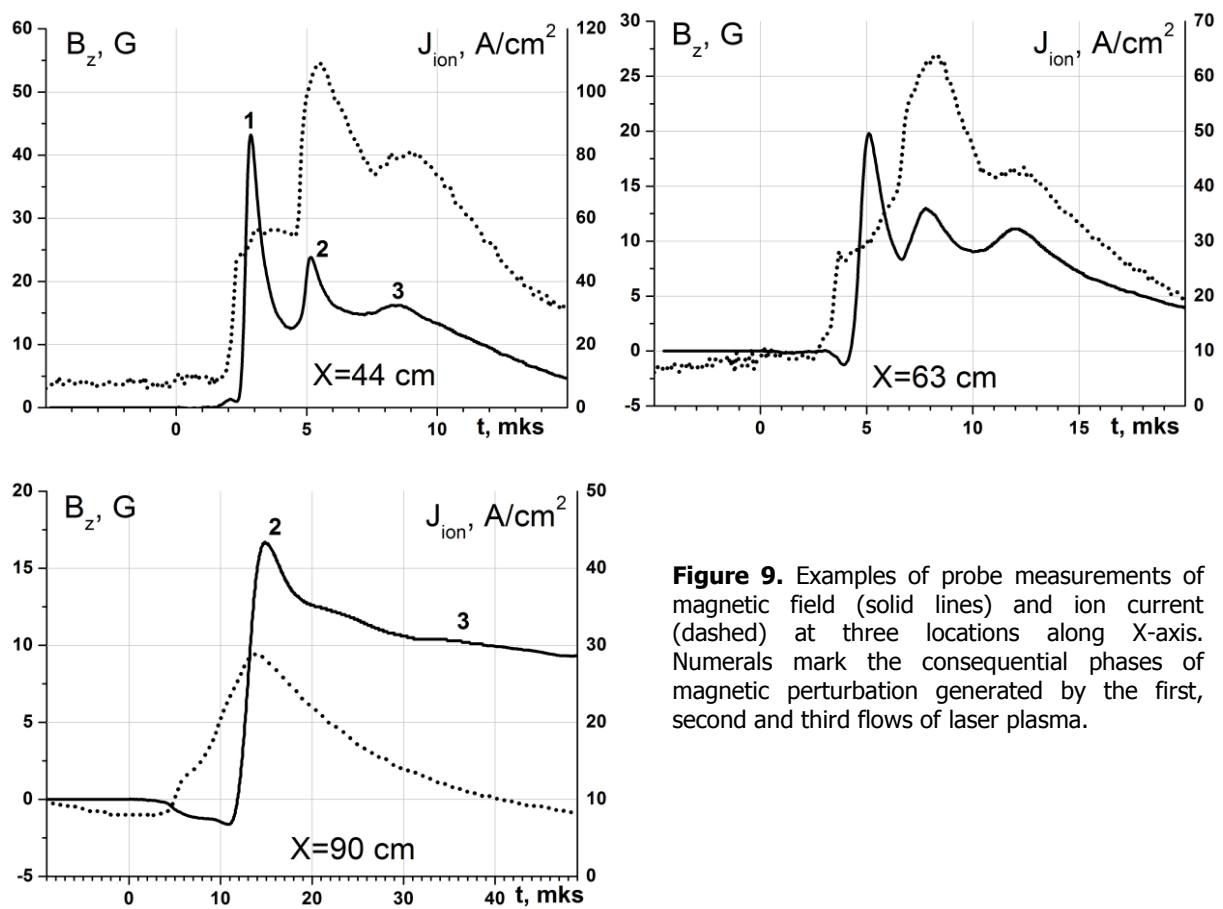

**Figure 9.** Examples of probe measurements of magnetic field (solid lines) and ion current (dashed) at three locations along X-axis. Numerals mark the consequential phases of magnetic perturbation generated by the first, second and third flows of laser plasma.

Fig. 9 demonstrates in more details the dynamic of magnetic field in question together with ion current at three different locations along X-axis. Each such measurement at a given position has been performed simultaneously during a single shot by probe set which combines Langmuir and magnetic probes. At location nearest to the dipole one can see that $B_z$ component closely follows the laser plasma. Three consequential LP flows carry local $B_z$ maximums. The first one is the largest, while the second and the third are of about the same amplitude. Signals at X=44 and 63 cm show that the first maximum of $B_z$ gradually falls behind the front of the first LP flow, while at farthest distance X=90 cm it disappears altogether. Such far from the target only $B_z$ associated with the second and third LP flows remain. Further on we will consider only the part of $B_z$ signal related to second and third LP flows which is much more permanent and durable feature than the first maximum. That the generated by LP magnetic field depends on the density of background demonstrates fig. 10 where dynamic signals of $B_z$ are shown for rarified and dense θ-pinch plasmas. It is seen that for rarified case the maximum field is carried by LP front. In denser background the magnetic field is significantly larger and is carried by the LP second flow, while the first LP flow doesn't carry magnetic field at all.

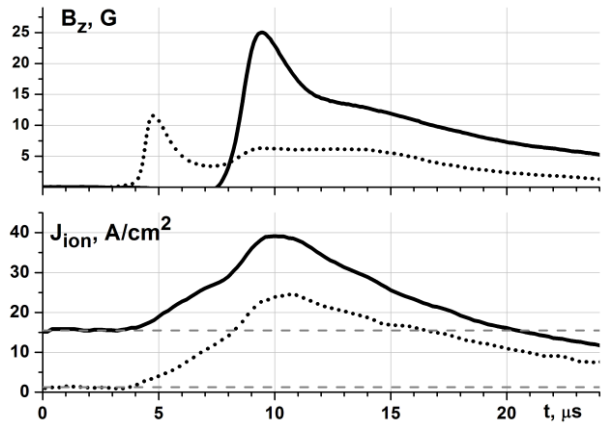

**Figure 10.** Dynamic signals of ion current (lower graph) and magnetic field (upper graph) measured at a distance of X=67 cm in rarified (dotted lines) and dense (solid lines) background plasma (levels prior to LP arrival are indicated by dashed horizontal lines).

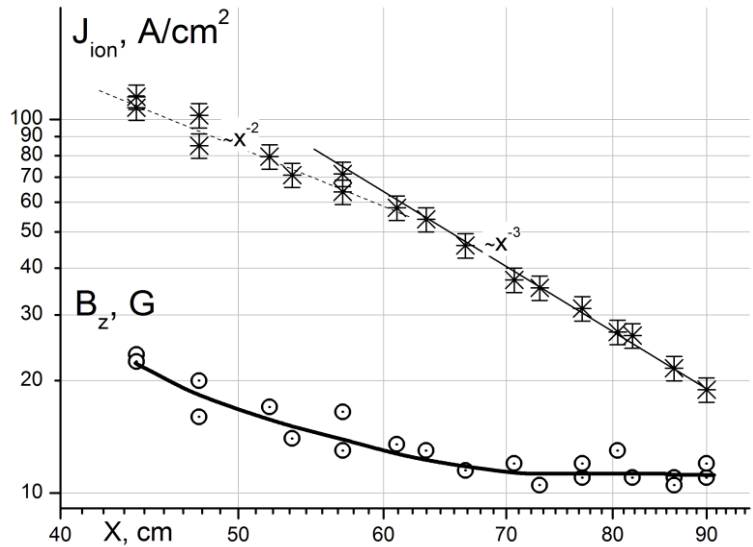

**Figure 11.** Spatial profiles of ion current (crosses) and magnetic field (circles) measured along the interaction axis. For the ion current amplitude the maximum of the second LP flow is taken. For the magnetic field the averaged value carried by the second and the third LP flows is calculated. Straight lines plotted over ion current data indicate square and cubic behavior.

By measuring dynamic signals at various locations the dependence of amplitude on the distance can be plotted. Fig. 11 shows such plots for the second maximum of ion current and carried by it magnetic field. One can see expected cubic decrease of LP flow with distance. At X<60 cm the fall off is more slow. This is probably explained by the two stream structure of LP consisting of two overlapping cones. The magnetic field shows relatively weak dependence on the distance from the target. At large X it levels off at a value of about 12 G. Note that at X=90 cm the dipole field value is less that 1 G. Without Θ-pinch plasma and in vacuum dipole field the amplitude of magnetic field carried by LP also varies little with distance and its level of about 2 G remains much smaller than in case when background plasma is present.

Measurements along Z axis revealed that $B_z$ component is broadly distributed above and below equatorial plane. Besides that, there were anti-symmetric (reversible) $B_y$ and $B_x$ components. They could be observed only when both background plasma and dipole field were present. Otherwise the values were below the resolution of magnetic measurements $\leq 1$ G. The sign of $B_x$ component corresponded to stretching of dipole field lines. The sign and value of $B_y$ component corresponded to positive electric current of the order of 1 A/cm$^2$ flowing along X axis. All three components of magnetic field are shown in fig. 12 for two positions along Z axis. For time reference the ion current is plotted as well.

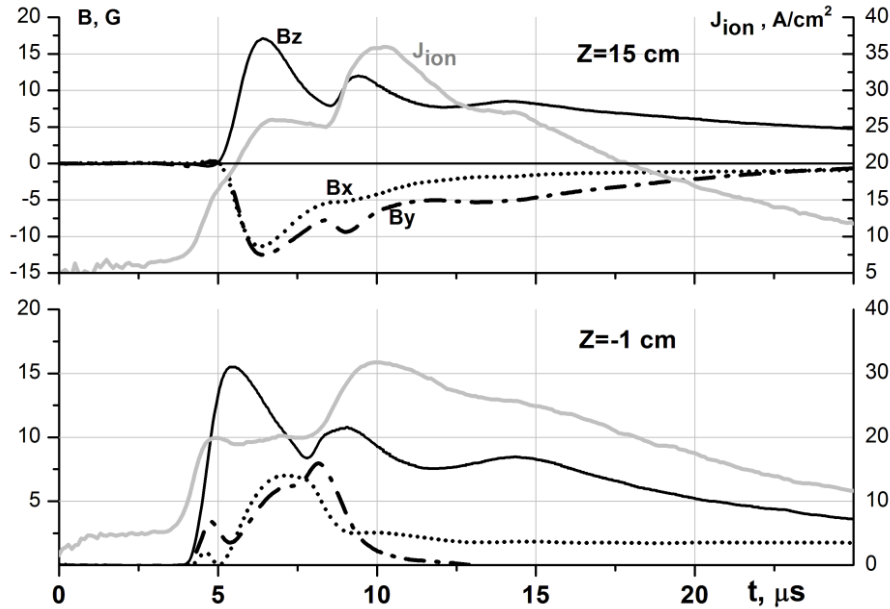

**Figure 12.** Dynamic signals of magnetic field components and ion current (grey lines) measured at a distance of X=67 cm and two different locations relative to the equatorial plane. Solid lines – $B_z$, dotted – $B_x$, dash-dot – $B_y$.

Upper panel reveals that reversible $B_x$ and $B_y$ components, like $B_z$, have a strong maximum associated with the first LP flow. The second and third LP flows carry relatively small $B_x$ and $B_y$ fields of the order of 5 G. Measurement close to equatorial plane (lower panel) shows that these components are practically zero at the time of passage of second and third LP flows.

### 3. Discussion and Conclusions

To access the conditions of experiment we list in the table 1 some of the most important parameters of Θ-pinch plasma, of magnetosphere created by it around the magnetic dipole and of laser plasma which is created inside of magnetosphere and expands into the background plasma. Before the laser plasma is produced on the surface of the dipole, the flow of background plasma creates around it a magnetosphere. While measurements close to the dipole were not conducted in this work, such data on laboratory magnetosphere are available from our previous studies (*Ponomarenko et al. 2004, 2008, Shaikhislamov et al. 2009*). In fact, in the present case the background plasma density of the order of $2 \cdot 10^{12}$ cm$^{-3}$ is sufficiently rarified that it should create a mini-magnetosphere, as was shown in a series of our dedicated experiments (*Shaikhislamov et al. 2013, 2014b, 2015b*). The so called Hall parameter, which is relation of pressure balance distance ($L_m = (\mu^2/2\pi n_* m V_*^2)^{1/6} \approx 30\,\text{cm}$) to ion inertia length ($c/\omega_{pi} \approx 15\,\text{cm}$), is equal to 2. For this value the structure of mini-magnetosphere differs from a typical planetary magnetosphere by partial penetration of plasma across magnetopause and inside of dipole dominated region. Based on the experiment with the same value of Hall parameter (*Shaikhislamov 2013*) it can be deduced that in present case the plasma should penetrate by as much as $L_m/2 \approx 15\,\text{cm}$. Thus, laser plasma when is created expands at first in the dipole magnetic field of inner magnetosphere filled from about X=15 to X=30 cm by background plasma and after that across magnetopause into the background plasma proper where magnetic field is zero. The value of magnetic field at magnetopause estimated from the pressure balance condition is about $B_m$=50 G. At the magnetopause the local Alfven speed is equal to the Θ-pinch velocity, that is about 50 km/s. Due to compression the magnetic field inside of magnetosphere behaves as $B = B_m/2 + \mu/X^3$ and the Alfven speed reaches 100 km/s at a distance of about X=20 cm. Considering now the laser plasma as it expands in magnetosphere, one can see that even the second laser plasma flow which moves with velocity of 100 km/s becomes super-Alfvenic at X>20 cm due to loading of dipole magnetic field with background.

**Table 1.** Parameters of experiment.

| Background plasma parameters | | |
|---|---|---|
| Velocity | $V_*$, km/s | 20-60 |
| Density[1] | $n_*$, cm$^{-3}$ | $2 \cdot 10^{12}$ |
| Mach number | $V_*/C_{s*}$ | ~1 |
| Magnetospheric parameters | | |
| Magnetosphere size[2] | $L_m$, cm | 30 |
| Relative dipole radius | $R_D/L_m$ | 0.1 |
| Knudsen number | $\lambda_*/L_m$ | ~2 |
| Reynolds number | $4\pi\sigma L_m V_*/c^2$ | ~100 |
| Hall parameter | $L_m \omega_{pi}/c$ | 2 |
| Degree of ion magnetization[3] | $L_m/R_{L*}$ | 3 |
| Laser plasma parameters | | |
| LP energy to magnetic energy[4] | $Q_{LP}/Q_d$ | ~1 |
| Velocity relative to background | $V_o/V_*$ | $\approx 3$ |
| Mean free path[5] | $\lambda_{i*}$, cm | $\approx 500$ |

[1] Evaluated close to the dipole location at X=0.
[2] Calculated as pressure balance distance between Θ-pinch plasma and dipole field pressures.
[3] Ion gyroradius is calculated for the value of magnetic field of $B_m$=50 G.
[4] $Q_{LP}$ is calculated as total energy of a single LP blob (16 J) divided by initial LP volume (~8 cm$^3$). $Q_{LP}$ is calculated as dipole field energy density at a distance of X=3.5 cm where the LP target is positioned.
[5] Coulomb scattering length of LP test ion in the Θ-pinch plasma at a density of 4 10$^{12}$ cm$^{-3}$.

Experimental results show that expanding laser plasma sweeps and partially expels background (fig. 5). This takes place due to Coulomb collisions. Because of strong dependence of cross-section on velocity the faster first LP flow interacts much weaker than the more slow second and third flows (fig. 7). The probability of Coulomb collisional scattering of background proton by the LP flow is equal to: $\upsilon_{pi}t \approx 2.8 \cdot 10^{12} Z_i^2 \int V_i^{-3} n_i dt$. The density over time integral can be expressed through total flux which depends on distance as: $\int V_i n_i dt \approx F_i/X^2$; $\upsilon_{pi}t \approx 1.8 \cdot 10^{13} V_i^{-4} F_i/X^2$. Here we took for LP ion charge a mean value of 2.5. Calculations using values from fig. 4, 6 yields for the first LP flow $V_i \approx 1.7 \cdot 10^7$ cm/s, $F_i = 6.7 \cdot 10^{17}$, $\upsilon_{pi}t \approx 1.4 \cdot 10^2/X^2$, and for the second – $V_i \approx 10^7$ cm/s, $F_i = 6.7 \cdot 10^{17}$, $\upsilon_{pi}t \approx 1.2 \cdot 10^3/X^2$. It follows that already at X=40 cm the probability to scatter background proton by the first LP flow is below 10%, while for the second it is significant 27% even at a distance of 67 cm. These estimates well agree with fig. 5.

Despite decrease of the amplitude of the second LP maximum in the range between X=100-300 cm induced by background the collisional interaction in general is rather weak and counter-streaming flows deeply penetrate into each other. As comparison of dynamic signals in fig. 7 shows the applied dipole field also doesn't significantly affect the laser plasma. According to the estimate in table 1, the energy of LP is comparable to that of dipole magnetic field already at the laser target and with LP expansion quickly becomes dominant. In vacuum LP carries with itself the magnetic field collinear to dipole magnetic field. When background plasma pre-fills the vacuum chamber this magnetic field is order of magnitude larger. It imposes a question of origin of such magnetic field carried by LP.

Described conditions principally differ from the previous experiments on LP expansion in magnetized uniform background in that the magnetized background in the present case is a compact localized shell. Beyond this magnetized shell the background plasma doesn't contain any magnetic field and the initially present dipole field is totally expelled as well. In a uniform magnetized background the super-Alfvenic LP expansion eventually decelerates and generates a non-linear magnetosonic wave (*Shaikhislamov et al. 2015a*) or, if strong enough, possibly a shock wave (*Zakharov et al. 2013*). In the present case LP expansion remains super-sonic for very long distances and without global magnetic field isn't decelerated by background plasma due to magnetic laminar mechanism or finite Larmor coupling, though is gradually scattered by Coulomb collisions. The novel feature is that LP picks up the magnetized compact shell and carries it along. The value of the carried magnetic field remains significant for distances much larger than the initial width of the shell.

Magnetic field is carried by electrons which move with ions to keep the plasma quasi-neutral. When LP density is larger than that of background the dynamics of magnetic field should closely follow LP passage. Exactly this is observed at distances X<70 cm. However, when background density becomes comparable or larger than the LP density the dynamics of magnetic field uncouples from LP and is governed by background. Namely because of this the first LP flow progressively with distance "looses" magnetic field until it is observed in time only when the second LP flow arrives at the point of measurement. It was observed also that the amplitude of the carried magnetic field increases with the density of background plasma (fig 10).

Let's consider the interaction in the frame of displaced electrons model developed for super-Alfvenic spherical expansion of explosive plasma into uniform magnetized background (*Longmire 1963, Wright 1971*). According to this model, when the electron density of expanding plasma significantly exceeds that of background, which is true in our case at distances X<30 cm where magnetized shell exists, the electrons of background together with frozen-in magnetic field are displaced and strongly compressed at the front of LP. The compression and strong increase of magnetic field at the LP front and its total expulsion inside of LP proper was measured in detail in our previous experiment with uniform background (*Shaikhislamov et al. 2015a*). However, in present experiment no such compression is seen. Measured fields don't exceed the expected values of >50 G in the initial magnetized shell. Moreover, observed magnetic field is present in the whole LP flow. It seems that in the process of LP sweeping over the magnetized shell LP electrons catch the magnetic field of the shell so it becomes frozen-in into LP itself. And the effectiveness of this process increases with density of background plasma initially present in the magnetized shell. It is seen even without background plasma, though the field caught and carried by LP in this case is much smaller. Most probably, without Θ-pinch the role of background plasma plays the foremost and rarified part of LP that cannot disrupt dipole field lines and fills them instead creating magnetized shell which later interacts with the main part of LP. The scenario of interaction inferred from the findings of the experiment is illustrated in the pictures of fig. 13.

On the base of the obtained experimental results a following general conclusion can be made. Plasma expanding outward from the inner region of magnetic dipole can interact with it by catching and dragging the magnetic field lines. The effectiveness of such process of transfer of magnetic field far from the dipole is directly related to the density of background plasma prefilling the magnetic field lines close to the dipole. Without pre-made plasma magnetized into dipole field lines the impulsive energetic plasma doesn't carry any significant field after crossing the dipole region. There can be two reasons why the LP catches and carries within itself the magnetized shell formed by background plasma. First, the dipole field lines loaded with plasma can't move faster than with the Alfven speed, and sufficiently fast impulsive flow can overcome the magnetized shell instead of displacing it. Second, the curvature of dipole field lines makes it possible for electrons of LP to mix with electrons of magnetized shell. Only by such mixing the LP might pick up the magnetic field instead of displacing it. The last feature is a main difference of the present work from previous studies of LP interaction with uniform magnetized background.

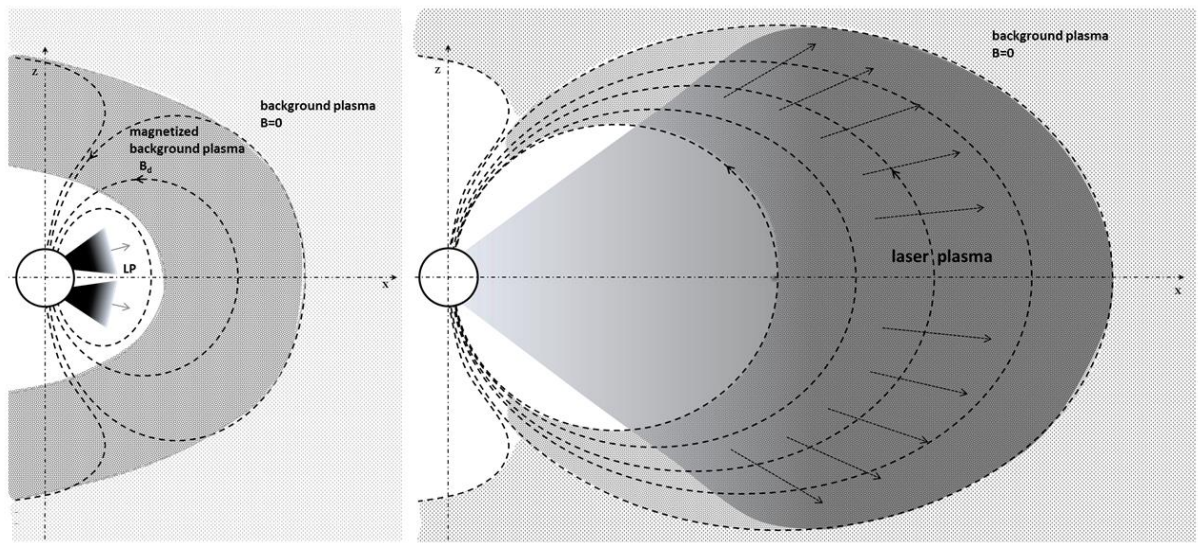

**Figure 13.** Schematic representation of the experimental results. Left picture – mini-magnetosphere including magnetized shell formed by Ө-pinch plasma around magnetic dipole. Right – laser plasma expanding flow which picks up and carries the magnetized shell. Dashed lines indicate magnetic field lines.

**Acknowledgements**
This work was supported by SB RAS Research Program (project II.10.1.4 N 01201374303) and Russian Fund for Basic Research grants 14-29-06036 and 16-52-14006.

**References**

Antonov V M, Bashurin V P, Golubev A I, Zhmailo V A, Zakharov Yu P, Orishich A M, Ponomarenko A G, Posukh V G and Snytnikov V N 1985 A study of the collisionless interaction of interpenetrating super-Alfven plasma flows *J. Appl. Mech. Tech. Phys.* **26** 757–763

Antonov V M, Boyarinsev E L, Boyko A A, Zakharov Y P, Melekhov A V, Ponomarenko A G, Posukh V G, Shaikhislamov I F, Khodachenko M L, Lammer H 2013 Inflation of a dipole field in laboratory experiments: Toward an Understanding of magnetodisk formation in the magnetosphere of a hot Jupiter *The Astrophysical Journal* **769** 28

Blanco-Cano X, Omidi N and Russell C T 2003 Hybrid simulations of solar wind interaction with magnetized asteroids: comparison with Galileo observations near Gaspra and Ida *J. Geophys. Res.* **108** 1216

Borovsky J E, Pongratz M B, Roussel-Dupre R A and Tan T-H 1984 The laboratory simulation of unmagnetized Supernova remnants: absence of a blast wave *Astrophys. J.* **280** 802–8

Cheung AY, Goforth R R and Koopman D W 1973 Magnetically induced collisionless coupling between counterstreaming laserproduced plasmas *Phys. Rev. Lett.* **31** 429–32

Funaki I, Kimura T, Ueno K, et al. 2007 in Proc. 30th Int. Electric Propulsion Conf., Florence, IEPC 2007–94, http://erps.spacegrant.org/uploads/images/images/iepc_articledownload_1988 2007/2007 index/IEPC-2007-094.pdf

Golubev A I, Solov'ev A A and Terekhin V A 1978 Collisionless dispersion of an ionized cloud into a homogeneous magnetized plasma *J. Appl. Mech. Tech. Phys.* **19** 602–9

Khazanov George, Peter Delamere, Konstantin Kabin and Linde T J 2005 Fundamentals of the plasma sail concept: Magnetohydrodynamic and kinetic studies *Journal of Propulsion and Power* **21** 853-861

Khodachenko M L, Shaikhislamov I F, Lammer H & Prokopov P A 2015 Atmosphere Expansion and Mass Loss of Close-orbit Giant Exoplanets Heated by Stellar XUV. II. Effects of Planetary Magnetic Field; Structuring of Inner Magnetosphere *The Astrophysical Journal* **813** 50

Longmire C L 1963 *Notes on Debris-Air magnetic interaction. Rand Corporation Report RM-3386-PR*

Moritaka Toseo, Hideyuki Usui, Masanori Nunami, Yoshihiro Kajimura, Masao Nakamura and Masaharu Matsumoto 2010 Full Particle-in-Cell Simulation Study on Magnetic Inflation Around a Magneto Plasma Sail *Plasma Science, IEEE Transactions* **38** 2219-2228

Omidi N, Blanco-Cano X, Russell C T, Karimabadi H and Acuna M 2002 Hybrid simulations of solar wind interaction with magnetized asteroids: general characteristics *J. Geophys. Res.* **107** 1487

Paul J W M, Daughney C C, Holmes L S Rumsby P T, Craig A D, Murray E L, Summers D D R and Beaulieu J 1971 Experimental study of collisionless shock waves No. CONF-710607--121; CN--28/J-9. United Kingdom Atomic Energy Authority, Abingdon (England). Culham Lab

Podgornyi I M and Sagdeev R Z 1970 Physics of interplanetary plasma and laboratory experiments *Physics—Uspekhi* **12** 445–462

Ponomarenko A G, Zakharov Yu P, Nakashima H, Antonov V M, Melekhov A V, Nikitin S A, Posukh V G, Shaikhislamov I F and Muranaka T 2001 Laboratory and computer simulations of the global magnetospheric effects caused by anti-asteroidal explosions at near-Earth space *Adv. Space Res.* **28** 1175–80

Ponomarenko A G, Antonov V M, Posukh V G, Melekhov A V, Boyarintsev E L, Afanasyev D M and Yurkov R N 2004 Simulation of non-stationary processes in the solar wind and its impact on the Earth magnetosphere *Report on MinPromNauka project Investigation of Solar activity and its magnifications in near Earth space and atmosphere* part III (in Russian)

Ponomarenko A G, Zakharov Yu P, Antonov V M, Boyarintsev E L, Melekhov A V, Posukh V G, Shaikhislamov I F and Vchivkov K V 2008 Simulation of strong magnetospheric disturbances in laser-produced plasma experiments *Plasma Phys. Control. Fusion* **50** 074015

Shaikhislamov I F, Antonov V M, Zakharov Yu P, Boyarintsev E L, Melekhov A V, Posukh V G and Ponomarenko A G 2009 Laboratory simulation of field aligned currents in an experiment on laser-produced plasma interacting with a magnetic dipole *Plasma Phys. Control. Fusion* **51** 105005

Shaikhislamov I F, Zakharov Y P, Posukh V G, Boyarintsev E L, Melekhov A V, Antonov V M and Ponomarenko A G 2011 Laboratory experiment on region-1 field-aligned current and its origin in the low-latitude boundary layer *Plasma Phys. Control. Fusion* **53** 035017

Shaikhislamov I F, Antonov V M, Zakharov Yu P, Boyarintsev E L, Melekhov A V, Posukh V G and Ponomarenko A G 2013 Mini-magnetosphere: Laboratory experiment, physical model and Hall MHD simulation *Adv. Space Res.* **52** 422

Shaikhislamov I F, Zakharov Yu P, Posukh V G, Melekhov A V, Antonov V M, Boyarintsev E L and Ponomarenko A G 2014a Laboratory model of magnetosphere created by strong plasma perturbation with frozen-in magnetic field Plasma *Plasma Phys. Control. Fusion* **56** 125007

Shaikhislamov I F, Zakharov Yu P, Posukh V G, Melekhov A V, Antonov V M, Boyarintsev E L and Ponomarenko A G 2014b Experimental study of a mini-magnetosphere *Plasma Phys. Control. Fusion* **56** 025004

Shaikhislamov I F, Khodachenko M L, Sasunov Y L, Lammer H, Kislyakova K G & Erkaev N V 2014c Atmosphere expansion and mass loss of close-orbit giant exoplanets heated by stellar XUV. I. Modeling of hydrodynamic escape of upper atmospheric material *The Astrophysical Journal* **795** 132

Shaikhislamov I F, Zakharov Yu P, Posukh V G, Melekhov A V, Boyarintsev E L, Ponomarenko A G and Terekhin V A 2015a Experimental study of collisionless super-Alfvénic interaction of interpenetrating plasma flows *Plasma Physics Reports* **41** 399-407

Shaikhislamov I F, Posukh V G, Melekhov A V, Zakharov Yu P, Boyarintsev E L and Ponomarenko A G 2015b West–east asymmetry of a mini-magnetosphere induced by Hall effects *Plasma Phys. Control. Fusion* **57** 075007

Slough J 2001 in Proc. 27th Int. Electric Propulsion Conf., Pasadena, CA, IEPC-01-202, http://www.ess.washington.edu/space/M2P2/iepc.slough.PDF

Winglee R M, Slough J, Ziemba T and Goodson A 2000 mini-magnetospheric plasma propulsion—tapping the energy of the solar wind for spacecraft propulsion *J. Geophys. Res.* **105** 21067–77

Winske D and Peter Gary S 2007 Hybrid simulations of debrisambient ion interactions in astrophysical explosions *J. Geophys. Res.: Space Phys.* **112**

Wright T P 1971 *Phys. Fluids* V. **14** p 1905

Zakharov Yu P, Ponomarenko A G, Antonov V M, Boyarintsev E L, Melekhov A V, Posukh V G and Shaikhislamov I F 2007 Laser plasma experiments to study super high-energy phenomena during extreme compression of the Earth's magnetosphere by Coronal Mass Ejections *IEEE PS* **35** 813–21

Zakharov Yu P, Antonov V M, Boyarintsev E L, Melekhov A V, Posukh V G, Shaikhislamov I F, Vchivkov K V, Nakashima H and Ponomarenko A G 2008 New Type of laser-plasma experiments to simulate an extreme and global impact of giant coronal mass ejections onto earth' magnetosphere *J. Phys: Conf. Ser. (Kobe, Japan)* **112** 042011

Zakharov YuP *et al* 2013 Large-scale laboratory simulation of space Collisionless Shocks in magnetized background by using Laser-Produced Plasma blobs of kJ-range effective energy *Proceed. VI Intern. Symp. on "Modern Problems of Laser Physics"* (*25–31 August 2013, Novosibirsk, Russia*) pp 193–4 (mplp.laser.nsc.ru)